\documentclass[twoside,reqno]{mbook}
\usepackage{epsfig,cite}
\usepackage{amssymb,amsmath}
\usepackage{times}
\begin{document}

\sloppy \raggedbottom

 \setcounter{page}{1}

\newpage
\setcounter{figure}{0}
\setcounter{equation}{0}
\setcounter{footnote}{0}
\setcounter{table}{0}
\setcounter{section}{0}



\title{Superscaling in lepton-nucleus scattering}

\runningheads{Superscaling in lepton-nucleus scattering}{M.B.Barbaro, 
J.E.Amaro, J.A.Caballero, T.W.Donnelly}

\begin{start}
\author{M.B.Barbaro}{1}, \coauthor{J.E.Amaro}{2}, \coauthor{J.A.Caballero}{3}, \coauthor{T.W.Donnelly}{4}

\address{Dipartimento di Fisica Teorica, University of Turin, and INFN, 
Sezione di Torino, 10125 Turin, ITALY}{1}
\address{Departamento de F\'\i sica At\'omica, Molecular y Nuclear,
Universidad de Granada, 18071 Granada, SPAIN }{2}
\address{Departamento de F\'\i sica At\'omica, Molecular y Nuclear,
Universidad de Sevilla, Apdo. 1065, 41080 Sevilla, SPAIN }{3}
\address{Center for Theoretical Physics, Laboratory for Nuclear Science
and Department of Physics,
Massachusetts Institute of Technology,
Cambridge, MA 02139, USA}{4}

\begin{Abstract}
We suggest that superscaling analyses of few-GeV inclusive electron scattering 
from nuclei, both in the quasielastic peak and in the region where the
$\Delta$-excitation dominates, allow one to make reliable predictions for
charge-changing neutrino reactions  at energies of a few GeV, relevant
for neutrino oscillation experiments. 
\end{Abstract}
\end{start}

\section[]{Introduction}

Neutrino scattering has been the object of several recent 
investigations in connection 
with ongoing and planned experiments exploring neutrino properties, 
which use nuclei as targets. 
Moreover, the availability of new neutrino beams and nuclear targets opens
the possibility of extracting information on the nucleon's properties,
in particular on its strange form factors.
For both purposes good control of the nuclear effects is essential 
for a reliable interpretation of the data.

Present analyses of neutrino oscillation experiments make use of the 
Relativistic Fermi Gas model, which, although reproducing the gross features
of inclusive electron scattering responses, does not account
for their detailed structure.
As will be shown in the following, relativitistic effects play an important 
role at the kinematical conditions relevant for modern experiments, namely 
neutrino energies of a few GeV: hence the necessity of more realistic 
nuclear modeling, which includes relativistic dynamics.

While a number of relativistic calculations for neutrino scattering have been
performed recently~\cite{semirel,PRL,Antonov_nu,Juan,Giusti,Co',Omar}, a different approach can
also be taken~\cite{PRC,NC}: since any reliable model for neutrino-nucleus cross sections 
should first be tested against electron scattering experimental data, one can 
try to extract the neutrino cross sections from the
experimental $(e,e')$ cross sections, where a large amount of data is 
available.

It has been shown~\cite{PRC,NC} that, in appropriate kinematical conditions, this
startegy can be pursued as a consequence of the superscaling properties of
the electron scattering data. 
We shall illustrate this procedure after a brief review of the formalism.

\section{Electron and neutrino scattering: formalism and 
Relativistic Fermi Gas}

Electron and neutrino scattering off complex nuclei are closely related
processes and as such they can be treated within the same formalism.

Apart from obvious differences in the kinematics, due to the 
leptonic masses involved in the two processes, the basic difference 
between $e$-$A$ and $\nu$-$A$ scattering is the nature of the exchanged
vector boson, a virtual photon, probing the electromagnetic nuclear current, 
in the former case and a $W^\pm$ or a $Z^0$, probing the weak current, 
in the latter.

As a consequence, the corresponding cross sections involve different 
nuclear response functions and can be in general written in a Rosenbluth-like 
form as:

\begin{eqnarray}
\left(\frac{d^2 \sigma }{d\Omega_e dk'_e }\right)_{(e,e')}^{(EM)}
 &=& \sigma _{Mott}
\left[{v}_{L}R_{L}+{v}_{T}R_{T}\right] 
\label{EMcs}
\\
\left(\frac{d^2 \sigma }{d\Omega_l dk'_l }\right)_{(\stackrel{(-)}{\nu},l^\mp)}^{(CC)}
&=& \sigma^{(CC)}_{0}
\left[\widehat{V}_{L}\widetilde R_{L}+\widehat{V}_{T}\widetilde R_{T}
\pm 
2 \widehat{V}_{T^{\prime }}\widetilde R_{T^{\prime }}\right]
\label{CCcs}
\\
\left(\frac{d^2 \sigma }{d\Omega_N dk_N }\right)_{(\stackrel{(-)}{\nu},N)}^{(NC)} 
&=& \sigma^{(NC)}_{0}
\left[ v_L \widetilde R_L + v_T \widetilde R_T + v_{TT} \widetilde R_{TT} 
\right.
\nonumber\\
&+&
\left.
v_{TL} \widetilde R_{TL} \pm \left(2 v_{T'} \widetilde R_{T'} +
2 v_{TL'} \widetilde R_{TL'}\right)
\right]
\label{NCcs}
\end{eqnarray}
for inclusive electromagnetic (in the extreme relativistic limit
$m_e\to 0$), weak charged-current 
($CC$, where the outgoing lepton $l$ is detected) and weak
neutral current ($NC$, where the nucleon $N$ is knocked out) reactions, 
respectively.
In the above $\sigma_{Mott}$ the Mott cross section and $\sigma_0^{(CC,NC)}$
the analogous quantities for the weak processes, $v_i$ and $\widehat V_i$
are factors depending upon the lepton kinematics (see Refs.\cite{PRC,NC} for 
their explicit expressions)
and the $\pm$ sign refers to neutrino or antineutrino scattering.

Note that whereas the $(e,e')$ cross section depends on two electromagnetic
respone functions, $R_{L,T}$, three weak responses, 
$\widetilde R_{L,T,T'}$, enter the $CC$ neutrino scattering due to the
presence of the axial weak current.
In the $NC$ reaction three more responses, $\widetilde R_{TT,TL,TL'}$ 
are involved, similarly to what happens in the semi-inclusive $(e,e'N)$ 
process.

Let us first focus on the quasielastic peak, where the dominant process is
the knockout of a single nucleon. This domain can be treated in first
approximation within the relativistic Fermi gas (RFG) model, where the nucleus is 
described as a collection of non-interacting Dirac particles.
In this case all the response functions in (\ref{EMcs},\ref{CCcs},\ref{NCcs}) turn out to 
be given by
\begin{equation}
(R_i)_{\rm RFG} = \frac{{\cal N} m_N}{q k_F}{R_i}^{s.n.} f_{\rm RFG}(\psi) ,
\label{respRFG}
\end{equation}
where $\cal N$ is the appropriate nucleon number, $m_N$ the nucleon mass,
$q$ the three-momentum transfer and $k_F$ the Fermi momentum. 
Moreover ${R_i}^{s.n.}$ are 
the corresponding single nucleon responses and 
\begin{equation}
f_{\rm RFG}(\psi) = \frac{3}{4} (1-\psi^2) \theta(1-\psi^2) 
\label{f}
\end{equation}
is the RFG {\em superscaling function}, which only depends upon a specific
combination of the variable $q$, $\omega$ (energy transfer) and $k_F$, namely
the {\em scaling variable}
\begin{equation}
\psi = \pm\sqrt{\frac{T_0}{T_F}}
\ ,\ \ \ \ \ \ \ 
T_0=\left(\frac{q}{2}\sqrt{1+1/\tau}-\omega/2
-m_N\right) .
\label{psi}
\end{equation}
In Eq.~(\ref{psi}) $T_F$ is the Fermi kinetic energy and $T_0$ the minimum
energy a nucleon in the Fermi sphere must have in order to participate in the
reaction at fixed $q$ and $\omega$, $\tau=(q^2-\omega^2)/(4 m_N^2)$ being the
dimensionless four-momentum transfer. The $\pm$ sign in (\ref{psi}) refers to 
values of $\omega$ smaller $(-)$ or larger $(+)$ than the quasielastic peak 
position $\omega=\sqrt{q^2+m_N^2}-m_N$, corresponding to $\psi=0$, and the
response region is limited by the $\theta$-function in (\ref{f}) to the range 
$-1\leq\psi\leq 1$. 

We stress that all the response functions, and hence the cross section, can
be factorized in a ``single-nucleon'' function times the superscaling function
$f$, which only depens upon $\psi$ and is the same for all nuclei: we then say
that the RFG model superscales, namely $f$ does not explicitly depend upon
the momentum transfer $q$ (scaling of the first kind) nor the Fermi momentum 
$k_F$ (scaling of the second kind).

In the $\Delta$-resonance peak, centered in 
$\omega=\sqrt{q^2+m_\Delta^2}-m_N$, all the above expressions still hold 
providing an appropriate scaling variable
\begin{equation}
\psi_\Delta = \pm\sqrt{\frac{T^\Delta_0}{T_F}}
\ ,\ \ \ \ \ \ \ 
T^\Delta_0=\left(\frac{q}{2}\sqrt{\rho+1/\tau}-\omega\rho/2
-m_N\right) ,
\label{psiD}
\end{equation}
accounting for inelasticity through the parameter 
$\rho=1+(m_\Delta^2-m_N^2)/(4\tau m_N^2)$, is used.
The RFG model superscales also in the $\Delta$ region
and the associated scaling function is the same as
in the quasielastic peak.

\section{The SuperScaling Analysis (SuSA) of electron scattering}

An extensive analysis of the world electron scattering data based
on the superscaling ideas introduced above has been performed in
the quasielastic domain~\cite{scaling} by studying a reduced cross section,
obtained as the ratio of
the experimental $(e,e')$ cross section and the appropriate 
single-nucleon factors, as a function of the scaling variable (\ref{psi}).

The analysis has shown that for high enough energies scaling of the first kind
is fulfilled at the left of the quasielastic peak (the so-called scaling 
region) and broken at its right,
whereas scaling of the second kind is very well satisfied at the left of the
peak and not so badly violated at its right. Moreover, the scaling violations
have been found to reside mainly in the transverse channel, whereas in the
longitudinal scaling is very good.

As a result a phenomenological superscaling function $f(\psi)$ has been
constructed which fits the electron scattering longitudinal 
response for all the available momentum transfers and nuclear species.
The deviations from the total (longitudinal and transverse) data can be 
ascribed to two contributions:
1) the excitation of the $\Delta$-resonance and 2) $q$- and $k_F$-dependent
nuclear effects such as meson exchange currents and their associated 
correlations, which 
break both kinds of scaling. Although the exact calculation of these effects
is very involved~\cite{MEC,DeltaMEC,Physrep,2p2h}, their net effect in the 
quasielastic region can be assessed to be roughly 10-15\%.

The superscaling function, plotted in Fig.~1 versus $\psi^\prime$ (the 
``prime'' indicates an energy
shift, of about 20 MeV, necessary to reproduce the right experimental
position of the quasielastic peak), significantly differs from the RFG one due
to nuclear interactions not included in the Fermi gas model. 
In particluar, it is
lower, it extends over a wider range of $\psi^\prime$ and, importantly, it
displays an asymmetric shape, with a pronounced tail in the $\psi^\prime>0$ 
region. This peculiar feature of the superscaling function, which has been
further investigated in recent work~\cite{PRL,Juan,Antonov_sc},
represents a stringent constraint on any reliable microscopic model.

\begin{figure}[h]
\centerline{\epsfig{file=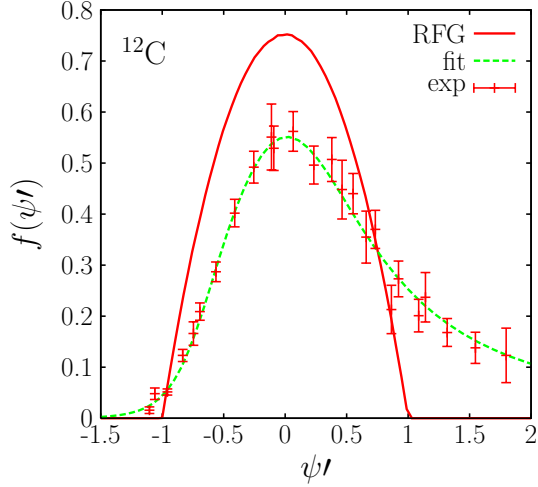,width=70mm}}
\caption{The experimental quasielastic superscaling function together with
its phenomenological fit (dashed green) and the RFG result
(solid red) plotted versus the corresponding scaling variable $\psi'$.
\label{fig1}}
\end{figure}

The same analysis has been recently extended to the $\Delta$ peak~\cite{PRC}
by first subtracting the quasielastic cross section calculated by means of the 
phenomenological superscaling function from the total experimental data
and then analyzing the result in terms of the scaling variable $\psi_\Delta$
defined in Eq.~(\ref{psiD}). This procedure
removes from the data the impulsive (i.e.,
elastic eN) contributions that arise from quasielastic
scattering. The corresponding reduced cross section (namely, divided by the
appropriate elementary $N\to\Delta$ function) has been shown to superscale
reasonably well for negative values of $\psi_\Delta$, indicating
that the dominant process contributing to the cross section in this region
has been correctly identified as the $N\to\Delta$ excitation. For positive
values of $\psi_\Delta$ this is no longer the dominant process since other
resonances and the tail of deep inelastic scattering come into play.

As a result a phenomenological superscaling function $f^\Delta(\psi_\Delta)$, fitting
the subtracted data, has been constructed in analogy with what was done in the 
quasielastic case (see Fig.~2). 

\begin{figure}[h]
\centerline{\epsfig{file=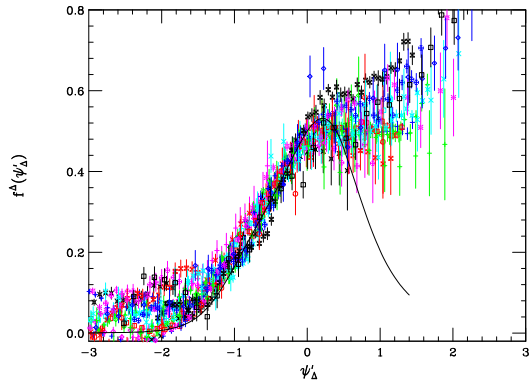,width=70mm}}
\caption{The experimental $\Delta$ superscaling function for the world data on
carbon and oxygen
and its phenomenological fit (solid line) 
plotted versus the corresponding scaling variable $\psi'_\Delta$.
\label{fig2}}
\end{figure}

A test of the two phenomenological superscaling functions is presented 
in Fig.~3,
where the data are compared with the result obtained by inserting these 
functions into the RFG responses (\ref{respRFG}) in place of $f_{\rm RFG}$.
In Ref.~\cite{PRC} the same test is performed at different kinematical conditions 
and for different nuclei, showing that the superscaling approach yields a very
satisfactory representation of the electron scattering data for a wide range
of kinematics, for energy transfer lower than the one corresponding to the 
$\Delta$ peak.

\begin{figure}[h]
\centerline{\epsfig{file=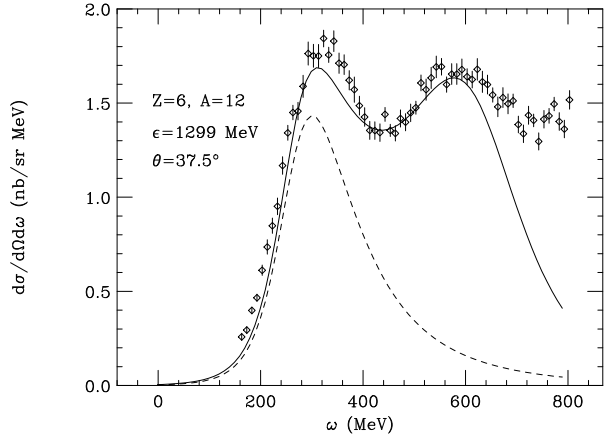,width=80mm}}
\caption{The experimental $(e,e')$ cross section for $^{12}C$ compared
with the SuSA prediction including both the quasielastic and $\Delta$ contribution (solid curve). The dashed curve represents the pure quasielastic cross section).
\label{fig3}}
\end{figure}

\section{Application of SuSA to neutrino and
antineutrino scattering}

Having succeded in representing the inclusive electron scattering 
data by means of
two ``universal'' (namely, $q$- and $A$-independent) superscaling functions,
valid in the quasielastic and $\Delta$-dominance regions respectively, we can
now reverse the procedure and give predictions for 
neutrino and antineutrino reactions.

Let us first focus on the charged-current process.
Since the kinematics of $(e,e')$ and $(\nu,\mu)$ are strictly related,
the procedure simply amounts to multiplying the phenomenological scaling function
by the appropriate elementary neutrino-nucleon factor in order to construct the
differential cross section (\ref{CCcs}).

The assumption underlying this procedure is that the susperscaling function
associated to the three responses in (\ref{CCcs}) is the same and is
equal to the one associated to the electromagnetic responses in (\ref{EMcs}).
While this is guaranteed by CVC as far as the vector-isovector components 
of the responses are concerned (i.e. the ones arising from the vector nuclear 
current), for the axial components it is not exactly true, since the
axial current is only partially conserved.
However, at the intermediate-to-high energies we are considering here, both
the axial and vector operators reduce, in the non-relativistic limit,
to $\sigma$~\footnote{Although we deal with relativistic effects exactly, this
argument can be made more clearly in the non-relativistic limit.}: hence the
corresponding scaling functions should be the same. This argument neglects
higher order contributions like meson exchange currents, which of course act
differently in the different channels, and go in any case beyond the present 
analysis.

In Fig.~4 we show the double differential quasielastic 
CC neutrino cross section with
respect to the outgoing muon momentum $k^\prime$ and solid angle $\Omega$
as a function of $k^\prime$, for neutrino energy of 1 GeV and muon scattering 
angle of 90 degrees.

It appears that the superscaling analysis prediction (solid red) is
significantly lower and wider that the RFG result (dashed green), 
reflecting the differences observed in Fig.~1.
For comparison the non-relativistic Fermi gas result (dotted violet) 
is also shown in order to stress the importance of relativistic effects in this
energy domain. Indeed these not only move the peak position and produce a
shrinking of the response region due to the relativistic kinematics, but they 
also yield a change in size due to the correct Dirac ``spinology''.
It clearly appears from Fig.~4 that relativistic effects cannot be neglected.
Whereas they can be accounted for exactly in the simple Fermi gas model, some
efficient prescriptions which can be used in order to ``relativize'' more
sophisticated nuclear calculations have been recently 
developed and successfully tested on the relativistic
shell model~\cite{semirel}.

\begin{figure}[h]
\centerline{\epsfig{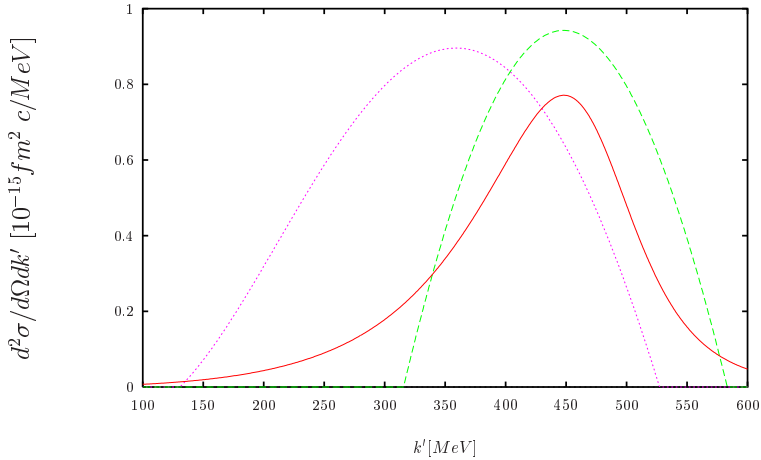}}
\caption{Double differential $(\nu_\mu,\mu^-)$ cross section versus
the outgoing muon momentum $k^\prime$ for neutrino energy $E_\nu=1$ GeV and
muon scattering angle $\theta_\mu=90^0$. 
Dashed (green): relativistic Fermi gas;
dotted (violet): non-relativistic Fermi gas; 
solid (red): SuSA result.
\label{fig4}}
\end{figure}

In Fig.~5 the neutrino and antineutrino differential cross sections are shown
for $E_\nu=1$ Gev and scattering angle of 45 degrees,
including also the $\Delta$ contribution (left peak). 
Note that the antineutrino cross section is about 5 times smaller than the
neutrino one, the difference becoming more dramatic as the scattering angle 
increases (see Ref.~\cite{PRC}). The reason is that the contributions of
the responses $\widetilde R_T$ and $\widetilde R_T^\prime$ to the cross section
are almost equal in size, but they add for neutrino and interfere 
destructively for antineutrino scattering. As a consequence small changes
in the model, as the inclusion of two-body currents, could have very large
effects on the $\overline\nu$ result, which should therefore be taken with
great caution.

\begin{figure}[h]
\centerline{\epsfig{file=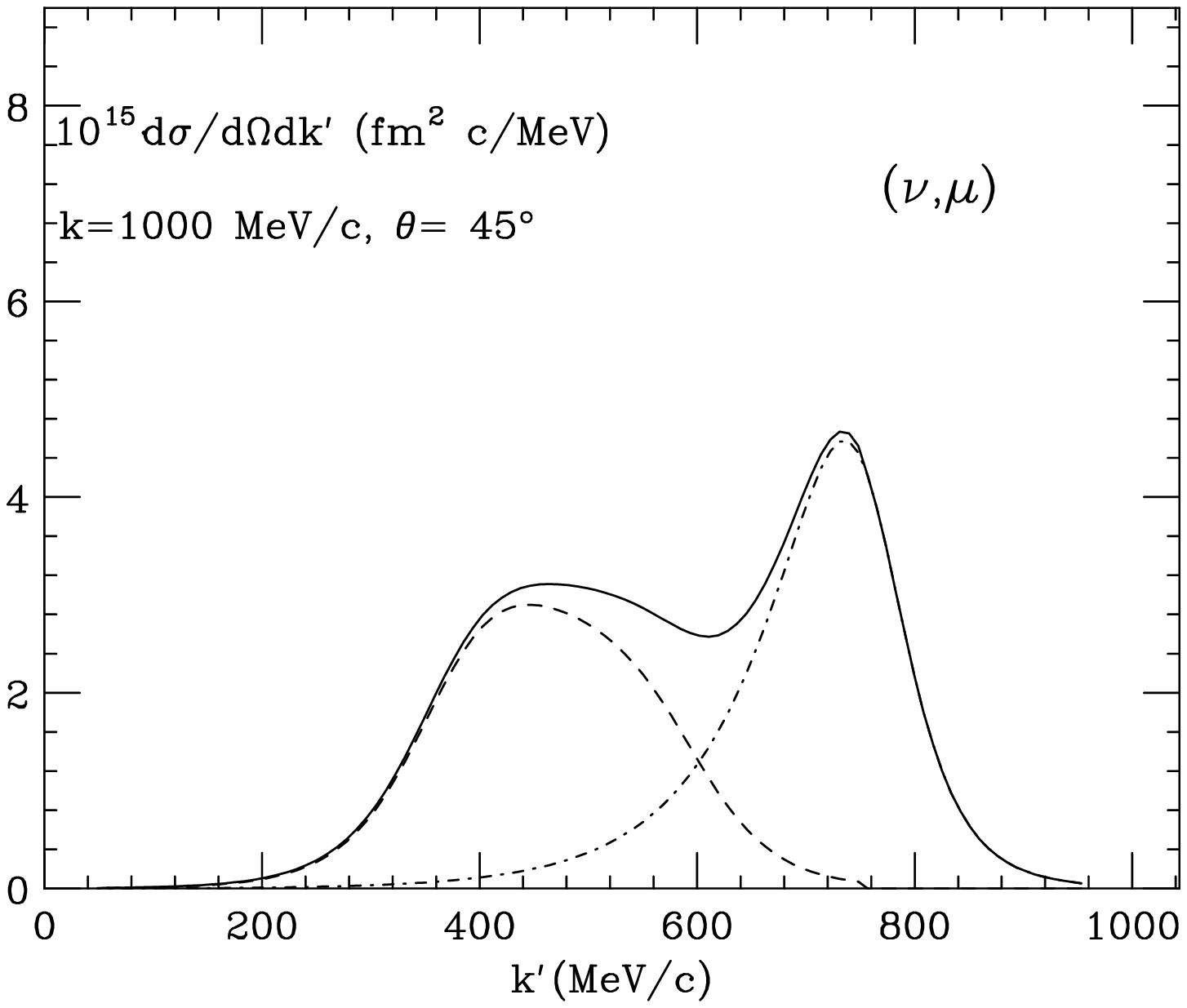,width=50mm}\epsfig{file=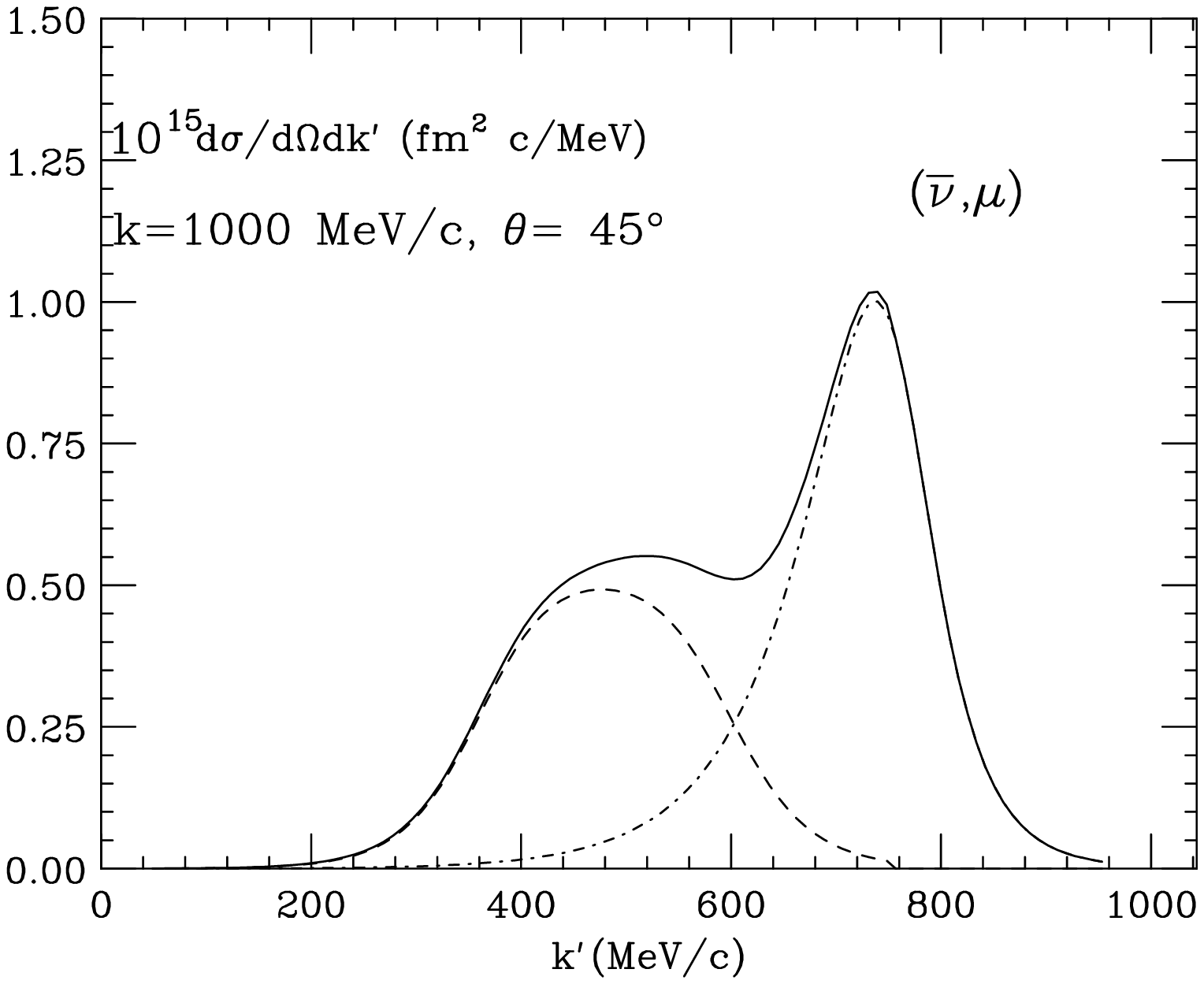,width=53mm}
}
\caption{
The SuSA cross section for the charged-current reactions 
$(\nu_\mu,\mu^-)$ (left panel) and $({\overline\nu_\mu},\mu^+)$
(right panel) on $^{12}$C  plotted
versus the final-state muon momentum $k'$. The 
dash-dotted curves give the QE contribution, the dashed curves the $\Delta$ 
contribution and the solid curves the total. 
\label{fig5}}
\end{figure}

In the case of neutral current reactions, the kinematics are different 
from the one of CC processes and of $(e,e^\prime)$.
In fact, while the CC reaction is a $t$-scattering type process 
(the Mandelstam variable $t$ is fixed), the NC is a $u$-scattering process,
since the detected final state is the outgoing nucleon, and the neutrino 
kinematic variables are integrated over~\cite{Mike}. This in turn implies
an integration region in the residual nucleus variables which is different in 
the two cases.
As a consequence, it is not obvious that the superscaling procedure, based
on the analogy with inclusive electron scattering, is still valid.

This issue is discussed at length in Ref.~\cite{NC}, where it is shown
that the scaling method is based on a factorization assumption which has been
tested numerically, with the outcome that the procedure can be applied
also to neutral current reactions.

In Fig.~6 (left panel) we show the results for quasielastic 
neutrino scattering and 
proton knockout calculated in both the RFG model (dashed green) and 
the SuSA approach (solid red). 
More results, corresponding to antineutrino 
scattering and to neutron knockout, are given in Ref.~\cite{NC}.

As for the CC case, it appears that the results based on superscaling are
significantly different from the ones of the non-interacting model and the
almost parabolic RFG cross section is strongly deformed by the nuclear
dynamics implicit in the phenomenological scaling function.

Finally, 
in Fig.6 (right panel) we show results obtained by allowing for a non-zero
strange quark content of the nucleon: we
compare our predictions for the cross section in a situation where no 
strangeness is
assumed (solid red) with the ones obtained including strangeness
in the magnetic (dashed green) and axial-vector (dotted violet) form
factors, using for $\mu_s=G_M^{(s)}(0)$ a representative value
extracted from the recent world studies of parity-violating electron
scattering and taking $g_A^s=G_A^{(s)}(0)$ to be
$-0.2$~\cite{PV}. The effects from inclusion of
electric strangeness are not shown here, since $G_E^{(s)}$ has
almost no influence on the full cross sections.

These results, which show a strong sensitivity of the cross section
to the strangeness content of the nucleon, are in line with the well-known 
fact (see, e.g., Refs.~\cite{Mike,Wanda}) that
the NC reactions can be used, together with parity-violating electron 
scattering~\cite{Musolf,Bill0}, 
to measure the strange form factors of the nucleon.

\begin{figure}[h]
\centerline{\epsfig{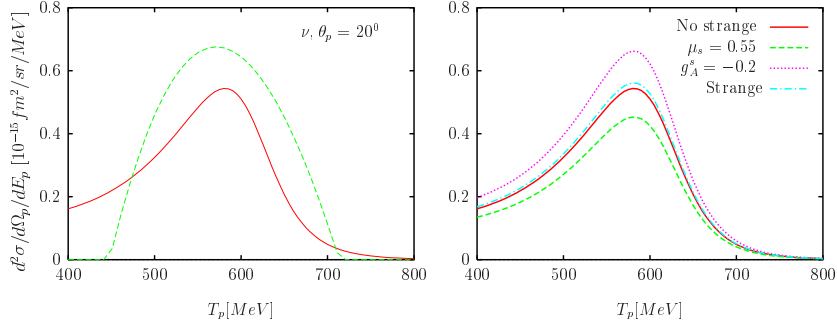}}
\caption{Double differential neutral current neutrino cross section
for proton knockout with respect to the proton energy and solid angle 
plotted versus the proton kinetic energy. 
The neutrino energy energy is 1 GeV and the proton scattering angle 20 degrees.
Left panel: the SuSA 
prediction (solid red) is compared to the RFG result (dashed 
green). Right panel: effects of variations of the magnetic 
(dashed green), axial (dotted violet) or both magnetic 
and axial (dotdashed cyan) strange form factors of the proton, compared with
the result without strangeness (solid red).
\label{fig6}}
\end{figure}

\section{Conclusions}

We have shown how the large amount of existing $(e,e')$ data
can be used to predict neutrino-nucleus cross sections, of interest for 
the search of neutrino oscillations and for the measurements of the
nucleon strange form factors.

The method we have developed is based on the superscaling properties displayed
by the electron scattering data both in the quasielastic peak and in the
$\Delta$-excitation region.

Beyond the practical usefulness of the SuSA procedure, 
this study opens several interesting questions connected to the microscopical
description of the superscaling function and in particular to its strongly
asymmetric shape, which represents a powerful test of different nuclear
models.

We believe that, for energies not too low, the superscaling analysis 
yields predictions for neutrino scattering at the 15-20\% level, the
uncertainty being related to the violations of superscaling.
These mainly arise from two body contributions
(meson-exchange currents and their associated correlations) which are not
included in our phenomenological representation of the nuclear dynamics.


\end{document}